\documentclass[twocolumn,showpacs,preprintnumbers,amsmath,amssymb]{revtex4-1}

\usepackage{graphicx}
\usepackage{dcolumn}
\usepackage{bm}

\begin{document}


\title{Radiation from a $D$-dimensional collision of shock waves: \\ a remarkably simple fit formula}

\author{Fl\'avio S. Coelho}
\author{Carlos Herdeiro}%
\author{Marco O. P. Sampaio}%
\affiliation{\vspace{2mm}Departamento de F\'\i sica da Universidade de Aveiro and I3N \\ 
Campus de Santiago, 3810-183 Aveiro, Portugal \vspace{1mm}}%

\date{March 2012}

\begin{abstract}
Recently \cite{Herdeiro:2011ck} we have estimated the energy radiated in the head-on collision of two equal $D$-dimensional Aichelburg-Sexl shock waves, for even $D$, by solving perturbatively, to first order, the Einstein equations in the future of the collision. Here, we report on the solution for the odd $D$ case. After finding the wave forms, we extract the estimated radiated energy for $D=5,7,9$ and $11$ and unveil a remarkably simple  pattern, given the complexity of the framework: (for all $D$) the estimated fraction of radiated energy matches the analytic expression $1/2-1/D$, within the numerical error (less than 0.1\%). Both this fit and the apparent horizon bound converge to $1/2$ as $D\rightarrow \infty$.
\end{abstract}

\maketitle

\noindent{\bf{\em I Introduction.}}
A collision of two point-like particles at trans-Planckian centre of mass energies is thought to form a black hole, with an associated emission of gravitational radiation \cite{Thorne,Choptuik:2009ww}. Moreover, as argued by t'Hooft  \cite{'tHooft:1987rb}, this process should be computable by classical General Relativity (GR). Understanding the physics involved is of conceptual interest, as a probe of the non-linear, non-perturbative, dynamical regime of GR,  and perhaps even of phenomenological interest, if the current or future generations of particle colliders, or of cosmic ray detectors, can probe trans-Planckian energies. This is a conceivable scenario if the fundamental Planck scale is much lower than the traditional four dimensional one, as suggested in the context of $D$ dimensional gravity (see \cite{Cardoso:2012qm}, Sec. 4, for a recent review).

One important property of this process is the inelasticity  $\epsilon$ of the collision: the percentage of the initial centre of mass energy radiated away. An upper bound for $\epsilon$ was provided in \cite{Eardley:2002re}  by an apparent horizon argument.
This method uses no information inside the future light cone of the collision and can be improved by computing the geometry therein. A method to do so which, albeit perturbative, carries information about a non-perturbative process, was devised by D'Eath and Payne \cite{D'Eath:1992hb,D'Eath:1992hd,D'Eath:1992qu}. This was recently extended to $D$ dimensions, to first order, by some of us  \cite{Herdeiro:2011ck}. Here, we report on the case of odd $D$. We find an extra contribution, not taken into account in~ \cite{Herdeiro:2011ck}  (and absent for even $D$), due to the different structure of the odd-$D$ Green's functions. Then, inspection of the result for $4\le D\le 11$, unveils a remarkable feature: the inelasticity obtained by the (technically and conceptually involved) D'Eath and Payne method matches, within the numerical precision of our method (error smaller than 0.1\%), the simple formula
\begin{equation}
\epsilon_{\rm 1st \,  order}= \frac{1}{2}-\frac{1}{D} \ . \label{miracle}
\end{equation}
Extrapolating this fit for large $D$ asymptotically matches the apparent horizon bound. 

The inelasticity obtained for  $4\le D\le 11$, together with the apparent horizon (AH) bound are shown in the following table (in percentage of centre of mass energy):
%
\begin{center}
\begin{tabular}{||  c || c | c | c | c | c | c | c |c |}
\hline			
  $D$ & 4 & 5 & 6 & 7 & 8 & 9 & 10 & 11\\
  \hline
  AH  bound & 29.3 & 33.5 & 36.1 & 37.9 & 39.3 & 40.4 & 41.2& 41.9\\
  \hline
  $\epsilon_{\rm 1st \,  order}$  & 25.0 & 30.0 & 33.3 & 35.7&37.5 & 38.9 & 40.0 & 40.9 \\
\hline  
\end{tabular}
\end{center}
%

\noindent{\bf{\em II Wave forms in odd dimensions.}}
The first order formalism of D'Eath and Payne was discussed in detail and generalised to higher $D$ in~\cite{Herdeiro:2011ck}. Two equal Aichelburg-Sexl shock waves collide head on in $D$ dimensions. The inelasticity of the process $\epsilon$ can be expressed as (we refer to~\cite{Herdeiro:2011ck} for all details) 
 \begin{equation}
\epsilon_{\rm 1st \,  order}=  \frac{1}{8}\dfrac{D-2}{D-3}\lim_{\hat{\theta}\rightarrow 0,r\rightarrow \infty}\left(\int (r\rho^{\frac{D-4}{2}}E_{,v})^2 dt \right) \; , \label{rad1}
\end{equation}
where the limit is selecting a radiation extraction point far away from the collision ($r\rightarrow \infty$) and along the collision axis ($\hat{\theta}\rightarrow 0$);  the wave form $E_{,v}$ at the space time-point $\mathcal{P}$ with null coordinates $u,v$ and at a distance $\rho$ from the symmetry (collision) axis is
\begin{eqnarray}
 E_{,v}&&(u,v,\rho)=-\dfrac{\sqrt{8}\Omega_{D-4}}{(2\pi u)^{\frac{D-2}{2}}} \int_{0}^{+\infty} \dfrac{d\rho'}{\rho'} \times \nonumber \\ &&
 \int_{-1}^{1} dx \,\dfrac{d}{dx}\left[x(1-x^2)^{\frac{D-3}{2}}\right] \delta^{(\frac{D-4}{2})}\left(\Delta v\right) \label{rad2}
 \; ,
\end{eqnarray}
where $\Delta v$ is selecting the events that support the radiation observed at $\mathcal{P}$.


\begin{figure*}
\includegraphics[scale=0.75,clip=true,trim= 0 0 0 0]{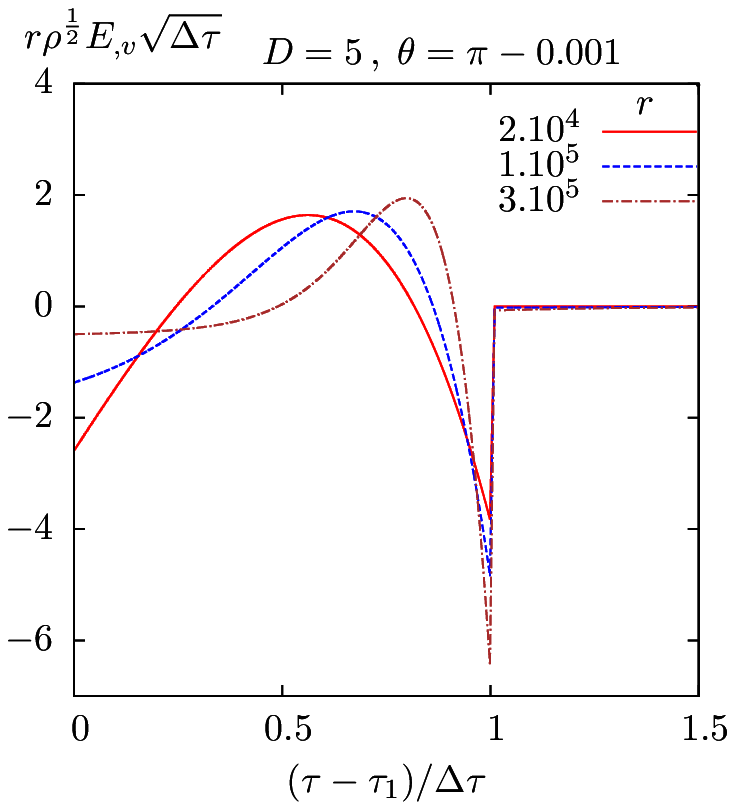}\hspace{3mm}
\includegraphics[scale=0.75,clip=true,trim= 0 0 0 0]{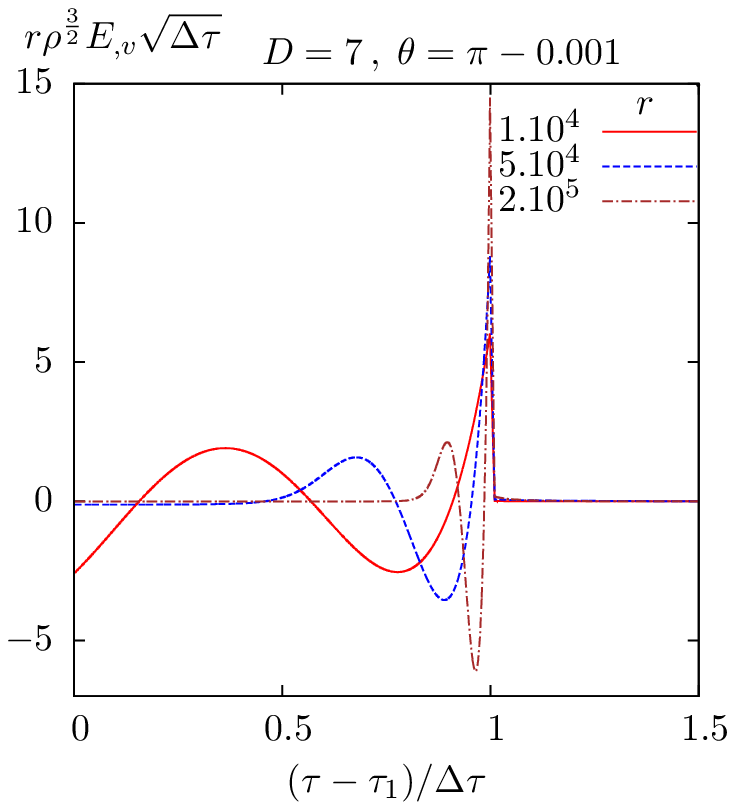}\hspace{3mm}
\includegraphics[scale=0.75,clip=true,trim= 0 0 0 0]{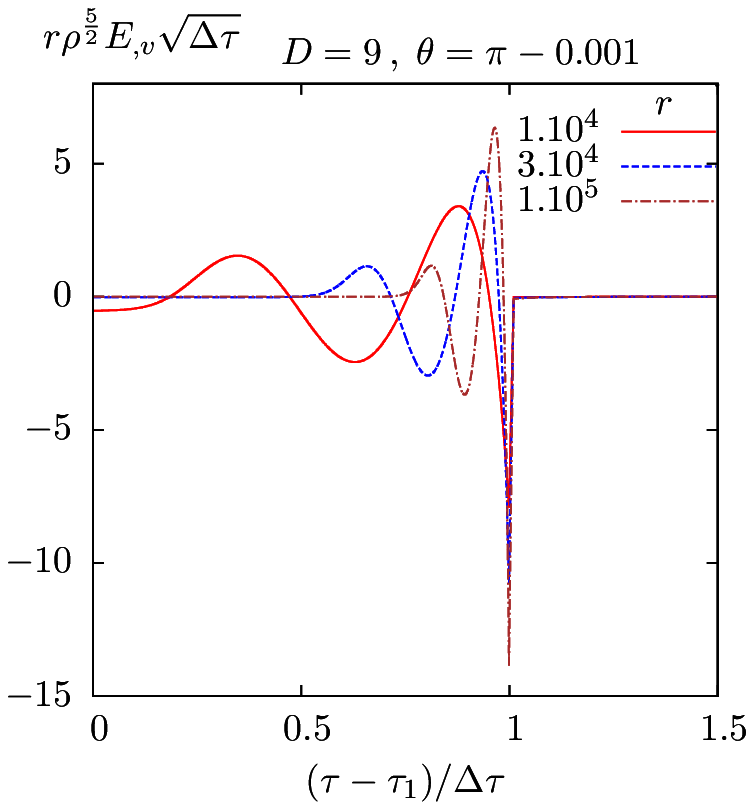}\hspace{3mm}
\caption{\label{fig:WaveForms}{\em Wave Forms for $D$ odd:} The panels contain wave form curves for the radiation signal seen by an observer close to the axis for various (large) $r$ as a function of time. The horizontal axis coordinate has been rescaled and shifted so that the times for the first and second optical rays coincide for the different curves.}
\end{figure*}

The main difference between the odd $D$ and even $D$ case in Eq.~\eqref{rad2} is the fractional derivative of the delta function denoted by its exponent. The fractional derivatives of delta functions have support not only at the zeros of their argument but also for positive argument. This is related to the well known property  that the retarded Green's function in odd dimensions has support not only on the past light cone, but also \textit{inside} the light cone \cite{friedlander}. In other words, odd dimensional flat spacetime behaves like a dispersive medium.  
 
 In~\cite{Herdeiro:2011ck}, the procedure followed for even $D$ was to perform $(D-4)/2$ integrations by parts over $x$ in Eq.~\eqref{rad2}, as to obtain a delta function and perform the $x$ integration completely. This procedure constrains the domain of integration of $\rho'$ to 
\begin{equation}
\mathcal{D}:\; -1 \leq x_\star \equiv \dfrac{U\Phi\left(\rho'\right)+\rho'^2-UT}{2\rho \rho'} \leq 1\; ,
\label{eq:domain}
\end{equation}
where $U = \tau + 2r\sin^2(\theta/2)$, $T=\tau +2r\cos^2(\theta/2)-\rho^2/U$ and $\tau,r,\theta$ are the usual retarded time, radial and polar angle coordinates (see~\cite{Herdeiro:2011ck}). For odd $D$ the procedure is similar, except that after $M=[(D-4)/2]$ integrations by parts there is a fractional delta function of order $1/2$, $\delta^{(1/2)}$. For this case, the $x$-integration of Eq. (B.14) of~\cite{Herdeiro:2011ck} is in fact more intricate. A careful integration by parts shows that the result is
\begin{equation}
r\rho^{M+\frac{1}{2}}E_{,v}=\dfrac{(-1)^M4\Omega_{D-4}}{(2\pi)^{M+2}(D-1)}\frac{r}{\rho} \int_{\mathcal{D'}} \dfrac{d\rho'}{\rho'^{M+\frac{5}{2}}}\dfrac{P^{(M+2)}(x_\star)}{\sqrt{1-x_\star}}
 \, , \label{odd}
\end{equation}
where now $\mathcal{D}'=\mathcal{D}\cup\mathcal{D}_{\rm extra}$, with $\mathcal{D}_{\rm extra}:\; x_\star \leq -1 \,$, and the polynomial $P^{(M+2)}(x_\star)$ factor in the domain $\mathcal{D}_{\rm extra}$ is replaced according to
\begin{equation}
\dfrac{P^{(M+2)}}{\sqrt{1-x_\star}}\equiv \sum_{k=0}^{M+2}\sum_{j=0}^k\dfrac{ c_k x_\star^{k-j}\left[\dfrac{(1-x_\star)^{j}}{\sqrt{1-x_\star}}-\dfrac{(-1-x_\star)^{j}}{\sqrt{-1-x_\star}}\right]}{(k-j)!j!(2j-1)} \; ,
\end{equation}
\begin{equation}
\dfrac{d^{M+2}}{dx^{M+2}}\left[(1-x^2)^{M+2}\right]\equiv \sum_{k=0}^{M+2}\dfrac{c_k}{k!} x^k \; .
\end{equation}
This extra term turns out to be crucial to obtain the correct late time tail of the wave forms.

A selection of wave forms is presented in Fig.~\ref{fig:WaveForms}. These were generated using the same numerical strategy as in~\cite{Herdeiro:2011ck}, with the extra term. We represent wave forms which have been rescaled by the relevant time scale for the problem, such that they start at retarded time $\tau_1$ and peak at $\tau_2$.  Such a time scale, $\Delta \tau=\tau_1-\tau_2$, is interpreted in the geometrical optics limit. For a (far away) observation point \textit{not} at the symmetry axis, a first ray arrives at the retarded time $\tau_1$ (corresponding to the beginning of the burst of radiation); then, a second ray arrives at $\tau_2$,  corresponding to the optical path that crosses a caustic at the axis before entering the curved region and hitting the observer (cf. Fig. 4 in~\cite{Herdeiro:2011ck}).

Observing the curves in Fig.~\ref{fig:WaveForms} one finds some similarities and differences with the even $D$ case. As for the similarities, the number of oscillations in the wave forms increases with $D$, with one more zero for each (from left to right in Fig.~\ref{fig:WaveForms}). Concerning the differences, the peak of radiation corresponding to the second optical ray is no longer singular, albeit becoming more pronounced as we increase $r$. The tails to the right of this peak are non-zero but integrable, since they decay as a power law. We have checked that the integrable tails are obtained from a cancellation of a non-integrable contribution from $\mathcal{D}$ and the contribution from $\mathcal{D}_{\rm extra}$. 

\begin{figure*}
\includegraphics[scale=0.73,clip=true,trim= 0 0 0 0]{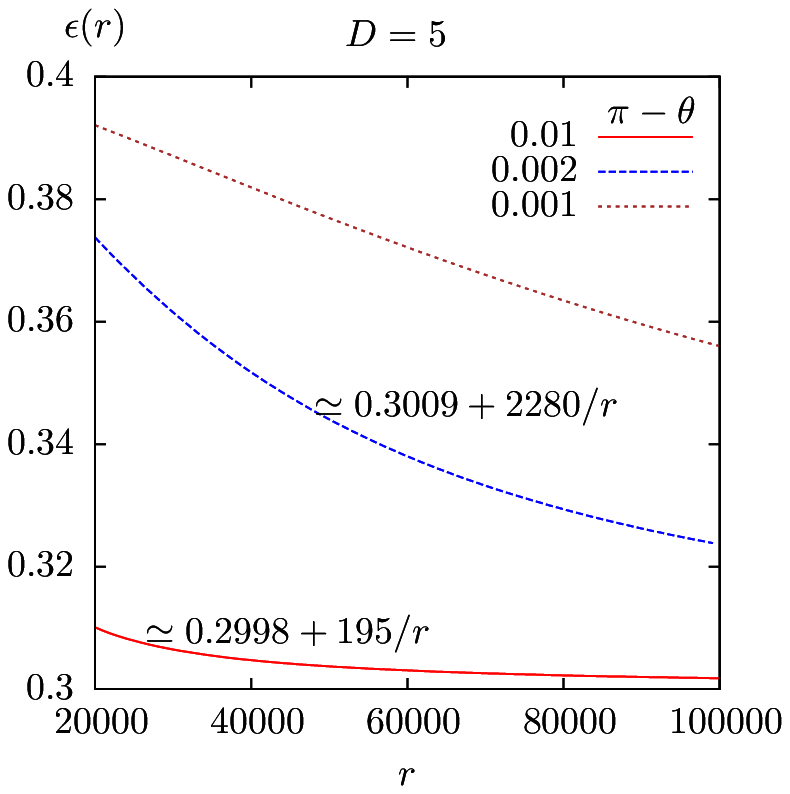}
\includegraphics[scale=0.73,clip=true,trim= 0 0 0 0]{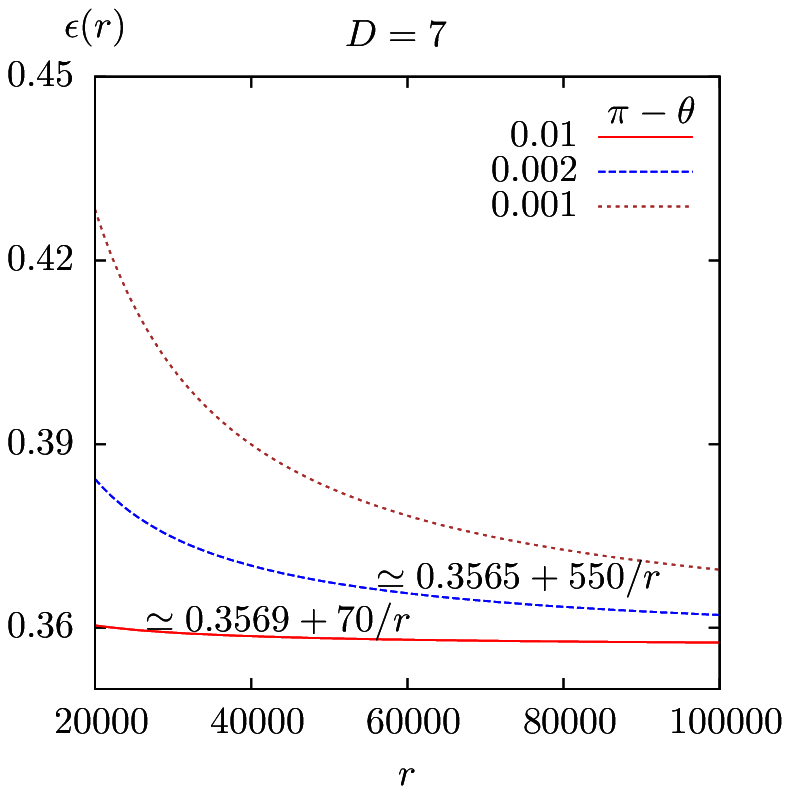}
\includegraphics[scale=0.73,clip=true,trim= 0 0 0 0]{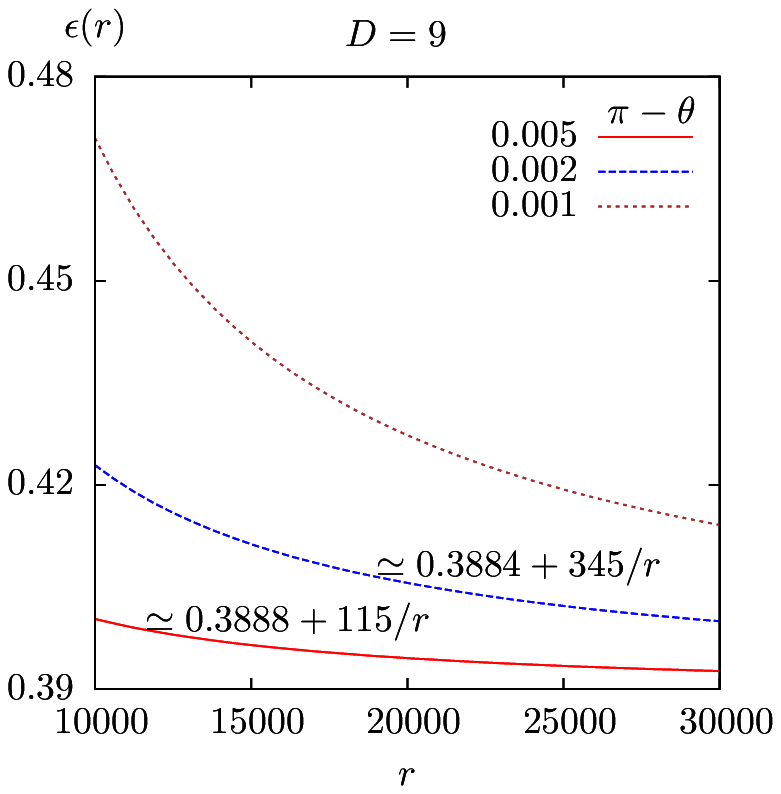}
\caption{\label{fig:ExtractRad}{\em Limiting fractions:} The panels contain, for each $D$, curves of $\epsilon(r)$ for some values  of $\theta$ which can be used to extract the limiting $\epsilon_{\rm 1st \, order}$. Some asymptotic curves which fit the numerical data to a high accuracy are indicated in each panel. The best estimates (relative error less than $10^{-3}$) are indicated by the constant terms in the asymptotic fits of the red solid curves.}
\end{figure*}
The inelasticity factor is  also extracted numerically, through the double limit in Eq.~\eqref{rad1}. For that purpose, we plotted the right hand side of Eq.~\eqref{rad1}, before taking the limit, as a function of $r$ in Fig.~\ref{fig:ExtractRad}, for several small $\hat\theta=\pi-\theta$ angles.  The most precise fit is extracted with $\hat \theta=0.01$. Similarly to even $D$, the result is extracted with a relative error smaller than $10^{-3}$.

\noindent{\bf{\em III Discussion.}}
The method presented in \cite{Herdeiro:2011ck} is technically involved, both analytically and numerically. It is quite reassuring that one can obtain the results for odd $D$ by the same method, fitting appropriately in the window bracketed by the neighbouring even $D$ values, even though they are obtained from integrating very different polynomial functions. This strongly legitimates the method we have used. 

\begin{figure}
\includegraphics[scale=0.8,clip=true,trim= 0 0 0 0]{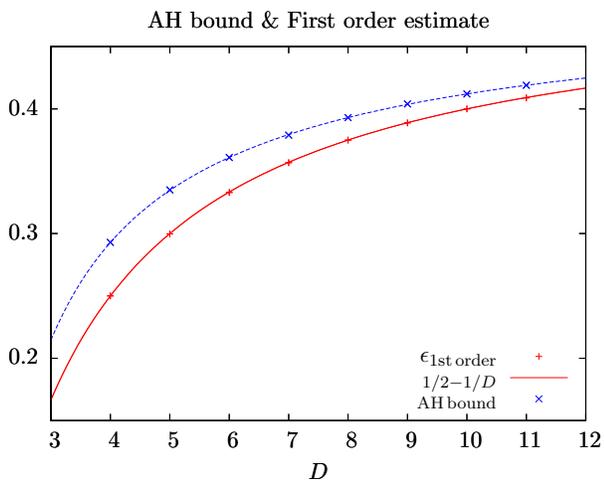}
\caption{\label{fig:ComparisonD} Apparent horizon bound (blue dashed line and points) compared with the first order result (red points) and our fit  - Eq. (\ref{miracle}) -  matching the result perfectly (red solid curve).}
\end{figure}

That the final result of this method fits, within numerical error (smaller than 0.1\%), with the simple formula given by Eq. (\ref{miracle}) is \textit{truly remarkable}. This agreement is exhibited in Fig.~\ref{fig:ComparisonD}, where the apparent horizon bound is also displayed.  It suggests the existence of a simpler physical or mathematical argument to derive the inelasticity in this process. This certainly deserves further investigation and motivates the study of higher order perturbation theory for this type of  processes. It is worth noting that, in second order perturbation theory, the matching conditions between the exact solution describing the two shock waves prior to the collision and the perturbative method are \textit{exact}. Moreover, in $D=4$, the second order result for the inelasticity ($16.3\%$ \cite{D'Eath:1992qu}) agrees with the value obtained in the high energy collision of two black holes in numerical relativity, within the numerical error ($14\pm3\%$ \cite{Sperhake:2008ga}).  Another suggestive fact is the convergence of both our fit and the apparent horizon bound to $1/2$ as $D\rightarrow \infty$, but its significance, or if it will hold in higher order perturbation theory, is yet to be unveiled.

\noindent{\bf{\em Acknowledgements.}}
 F.C. and M.S.  are funded by FCT through the grants SFRH/BD/60272/2009 and SFRH/BPD/69971/2010. The work in this paper is also supported by the grants CERN/FP/116341/2010, PTDC/FIS/098962/2008,  PTDC/FIS/098025/2008 and  NRHEP--295189-FP7-PEOPLE-2011-IRSES.

\bibliographystyle{h-physrev4}
\bibliography{shocks}


\end{document}